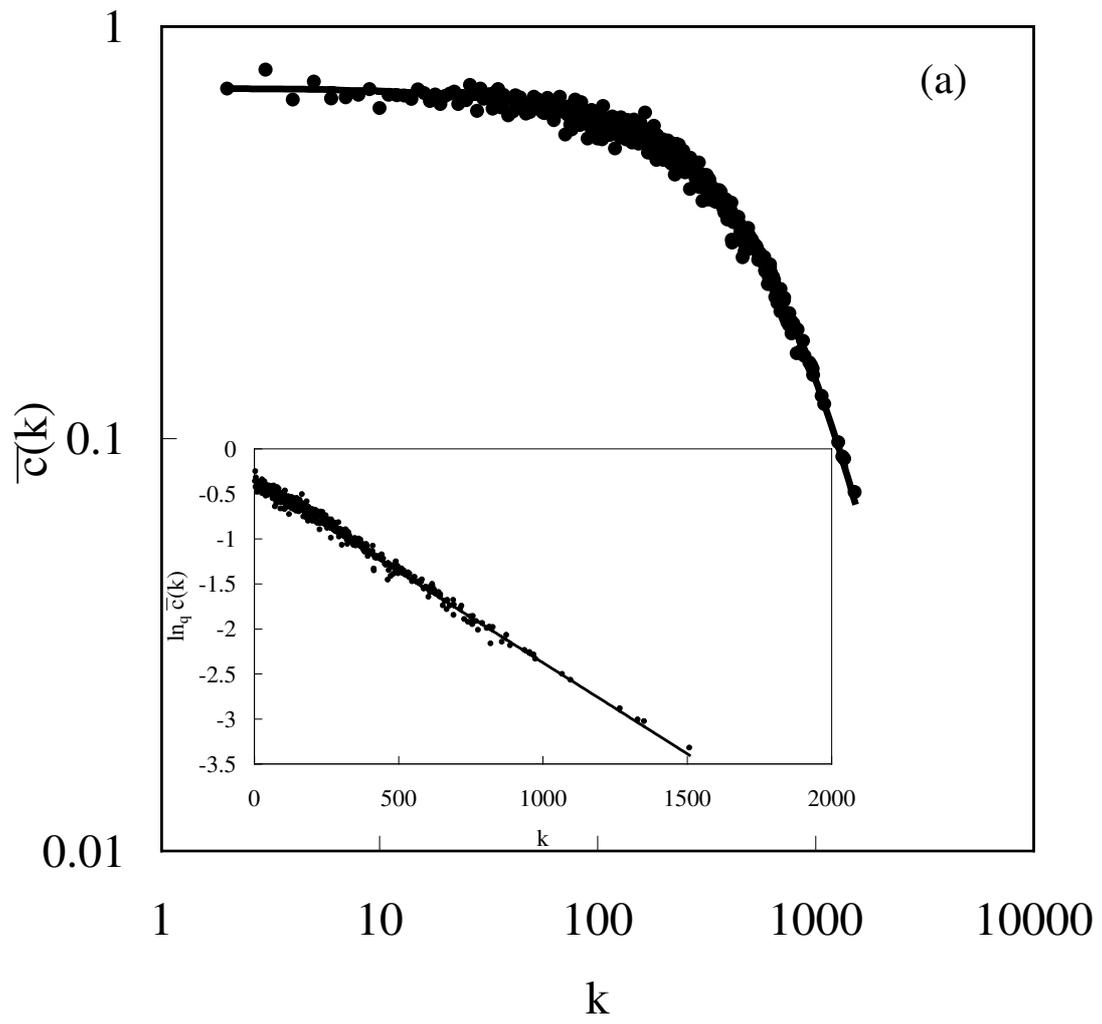

FIG. 1

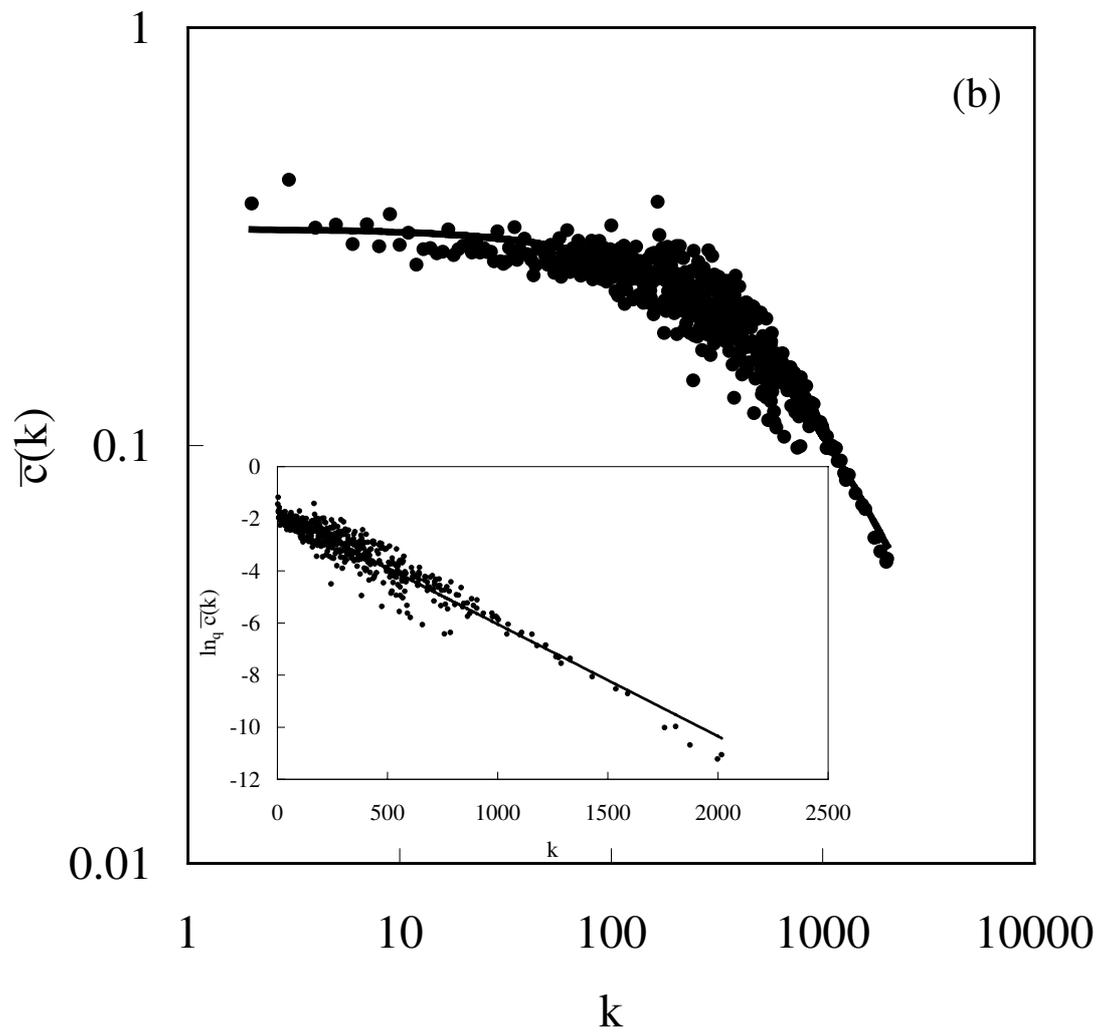

FIG. 1

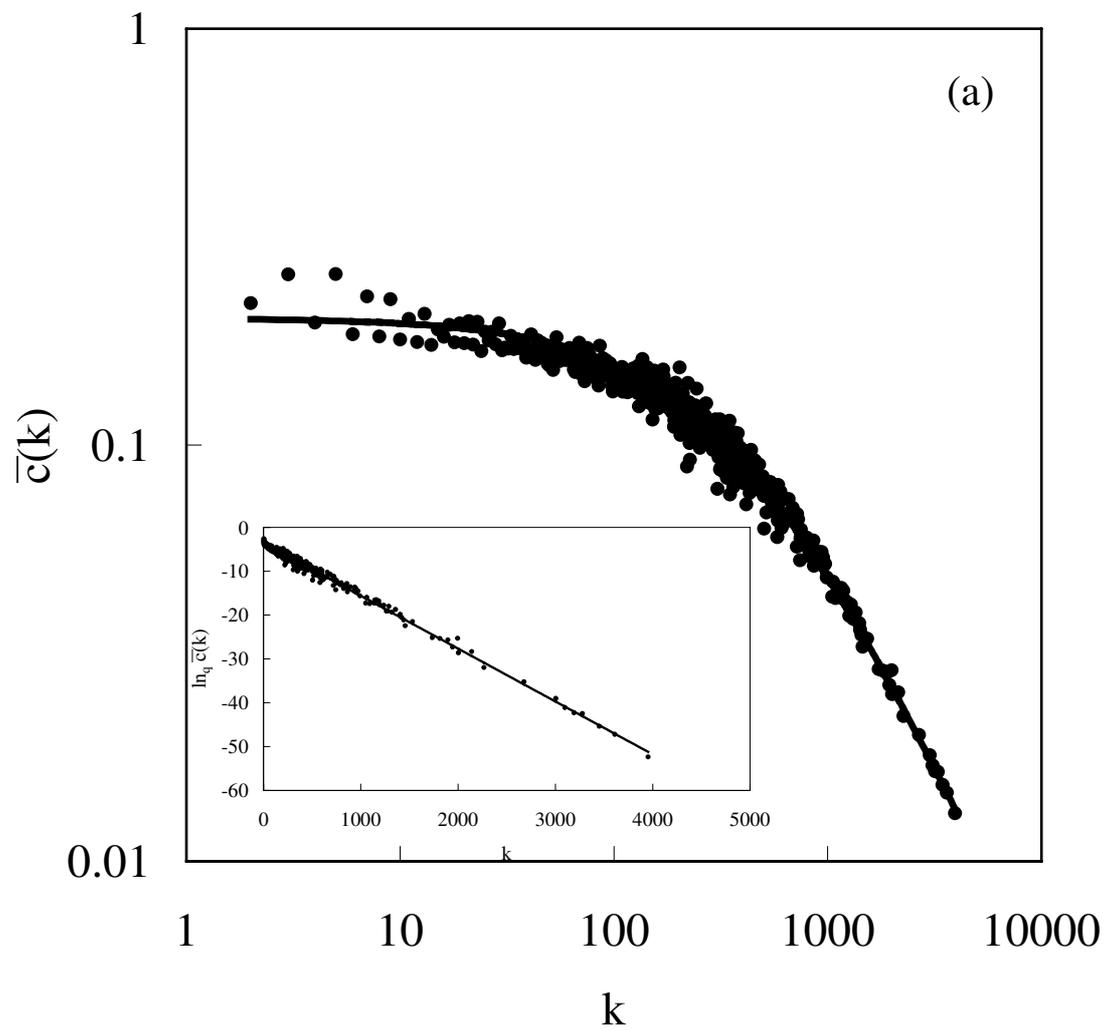

FIG. 2

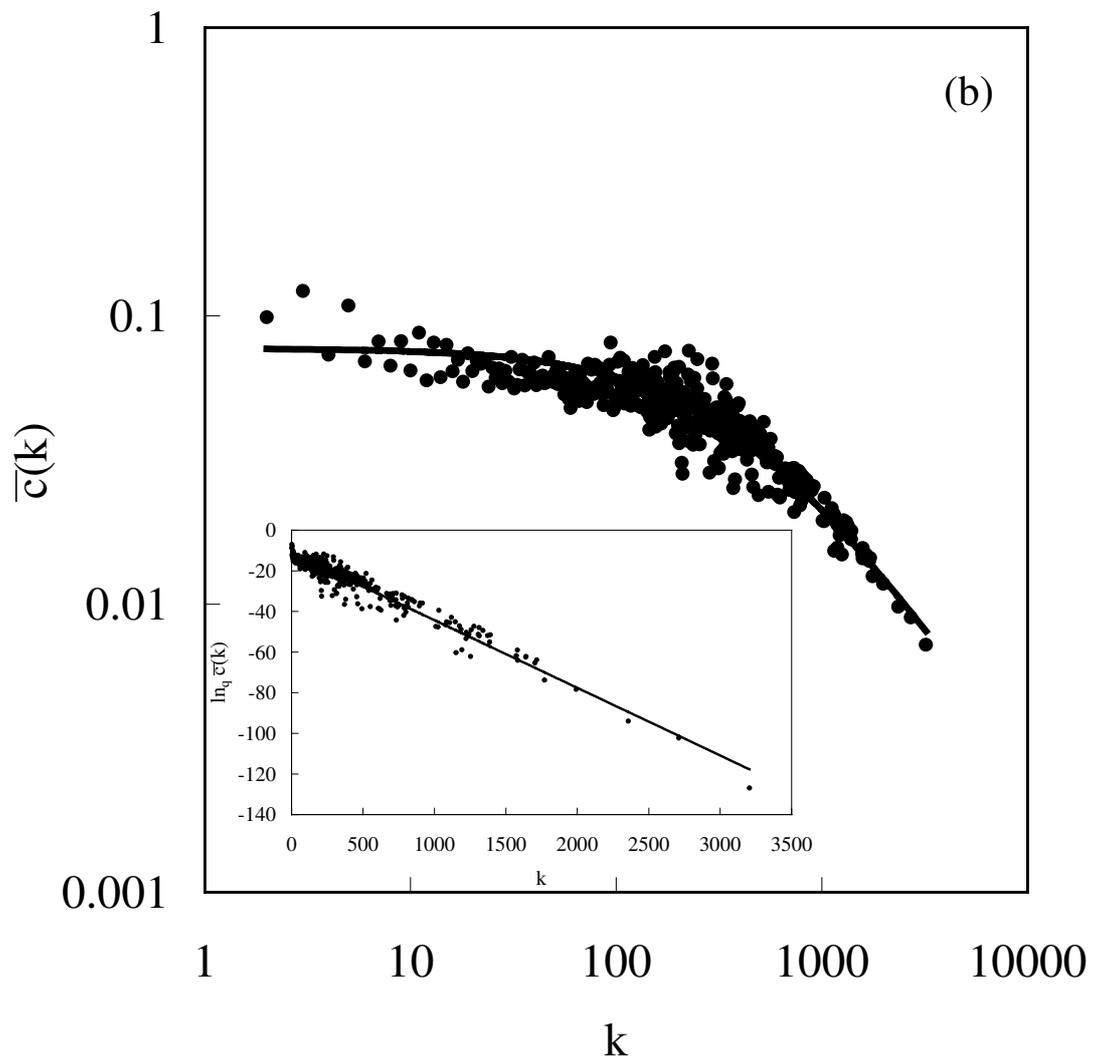

FIG. 2

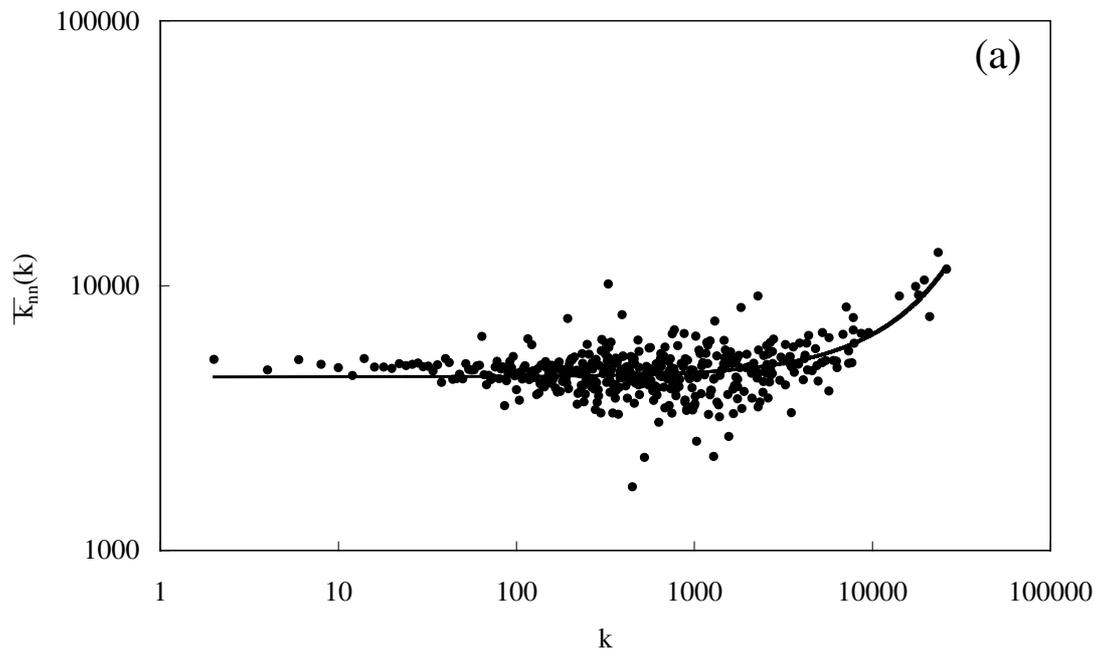

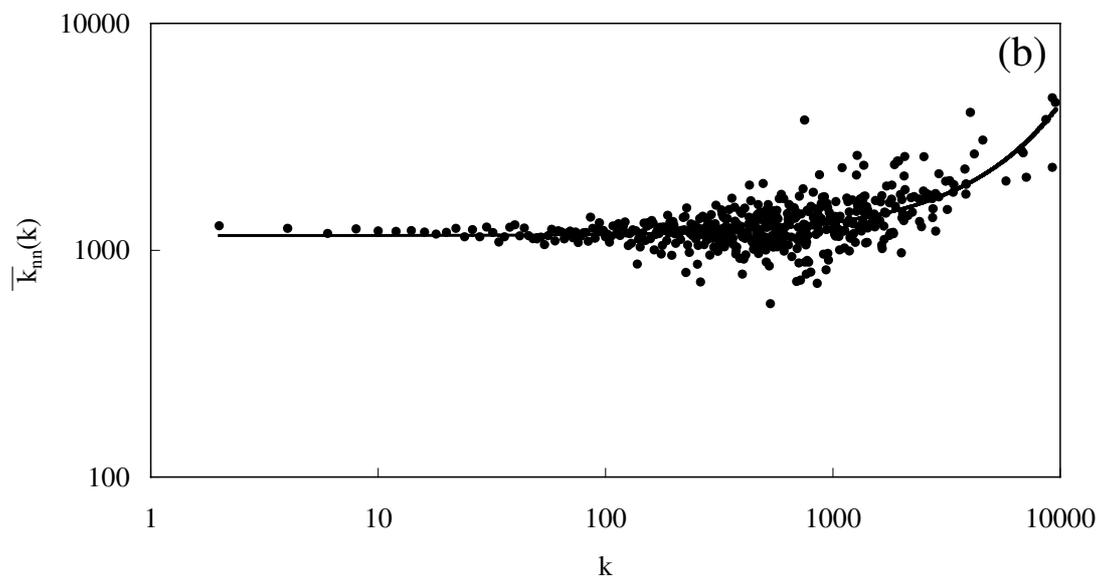

FIG. 3

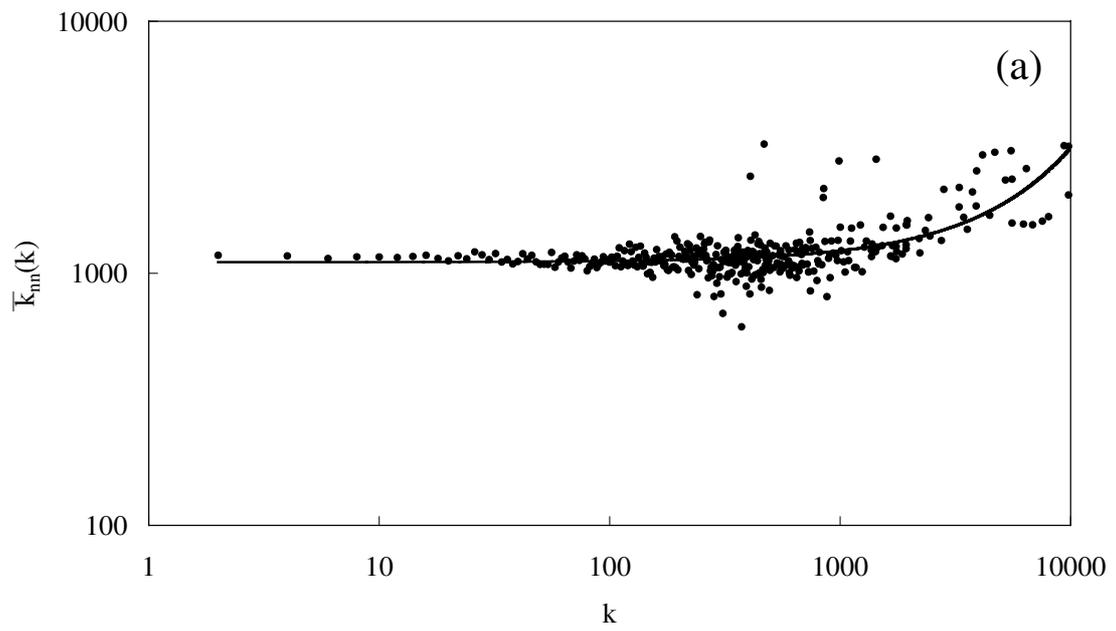

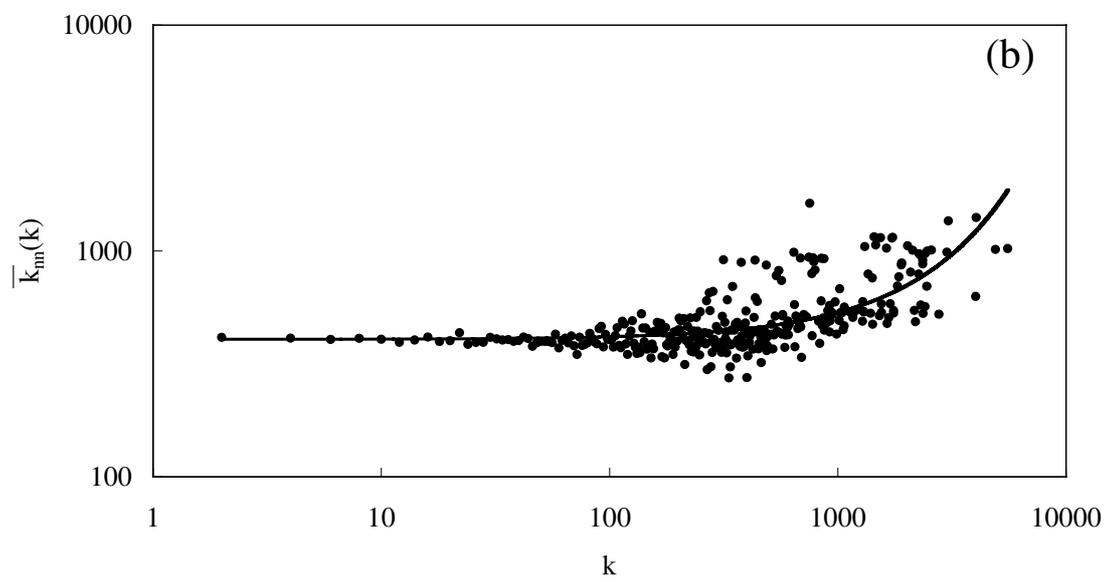

FIG. 4

| Year | 1984 | 1984-86 | 1984-94 | 1984-2004 |
|---|---|---|---|---|
| $N$ | 1200 | 1956 | 2999 | 3913 |
| $E$ | 18090 | 53993 | 208915 | 379726 |
| $<k>$ | 30.152 | 55.209 | 139.32 | 194.08 |
| $<c>$ | 0.388 | 0.476 | 0.577 | 0.635 |
| $<l>$ | 2.69 | 2.63 | 2.53 | 2.52 |

TABLE I

|  | 10km×10km×10km | 5km×5km×5km |
|---|---|---|
| California | $r = 0.285$ | $r = 0.268$ |
| Japan | $r = 0.161$ | $r = 0.156$ |

TABLE II

# Complex earthquake networks: Hierarchical organization and assortative mixing

Sumiyoshi Abe[1,*] and Norikazu Suzuki[2,**]

[1] *Institute of Physics, University of Tsukuba, Ibaraki 305-8571, Japan*

[2] *College of Science and Technology, Nihon University, Chiba 274-8501, Japan*

**Abstract** To characterize the dynamical features of seismicity as a complex phenomenon, the seismic data is mapped to a growing random graph, which is a small-world scale-free network. Here, hierarchical and mixing properties of such a network are studied. The clustering coefficient is found to exhibit asymptotic power-law decay with respect to connectivity, showing hierarchical organization. This structure is supported by not only main shocks but also small shocks, and may have its origin in the combined effect of vertex fitness and deactivation by stress release at faults. The nearest-neighbor average connectivity and the Pearson correlation coefficient are also calculated. It is found that the earthquake network has assortative mixing. This is a main difference of the earthquake network from the Internet with disassortative mixing. Physical implications of these results are discussed.

───────────────────────────────

*suabe@sf6.so-net.ne.jp

**suzu@phys.ge.cst.nihon-u.ac.jp

Earthquake phenomenon has been attracting continuous interest of physicists from the viewpoint of science of complex systems. In particular, seismicity exhibits several remarkable features, such as the scaling relation between frequency and magnitude called the Gutenberg-Richter law [1], and slow decay of the rate of aftershocks known as the Omori law [2]. In the recent works [3,4], we have analyzed the spatio-temporal properties of seismicity with the help of nonextensive statistics [5], which is a generalization of Boltzmann-Gibbs statistics and offers a consistent framework for treating complex systems. We have found that both the spatial distance and time interval between two successive earthquakes are nicely described by the $q$-exponential distributions, which are characteristics of nonextensive statistics and maximize the Tsallis entropy [6] under appropriate constraints. The fact that two successive earthquakes obey such definite statistical laws means that successive events are strongly correlated, no matter how large their spatial distance is. In fact, there is an investigation [7], which claims that an earthquake can be triggered by a foregoing earthquake that is more than 1000 km away. This implies that the seismic correlation length is enormously large, exhibiting a similarity between seismicity and phase transition phenomena. Thus, it seems inappropriate to put spatial windows in analysis of seismicity, in general.

In contemporary science, much attention is paid to statistical mechanics of complex networks [8-10], which provides a novel procedure for analysis of man-made as well as natural complex systems. In the network picture, vertices and edges represent elements and interrelation (i.e., interaction or correlation) between them, respectively. A primary purpose there is to understand the topological, statistical, and dynamical features of the networks.

In recent works [11-13], the concept of complex networks has been introduced to seismology to reveal spatio-temporal complexity of seismicity. A network associated with seismicity is constructed as follows. The geographical region under consideration is divided into small cubic cells with a certain size. (Since there are no *a priori* principles to determine the size of the cell, it is essential to examine the dependencies of the obtained results on it.) If events with any values of magnitude occurred in a cell, such a cell is identified with a vertex. As mentioned above, two successive events are supposed to be highly correlated, irrespective of their spatial distance. Such correlation is represented by an edge, here. Two vertices may coincide with each other (i.e., successive events occurring in the same cell), forming a loop. In this way, the seismic data is mapped to a growing random graph, termed the earthquake network. (Another method of constructing a network for earthquakes, which is much more complicated than the present one, is found in Refs. [14,15].) The earthquake networks thus constructed for the data taken in California and Japan have been analyzed in Refs. [11-13]. There, it has been discovered that they are small-world and scale-free networks. Emergence of scaling is due to the empirical fact that *aftershocks associated with a main shock tend to return to the locus of the main shock geographically*. The stronger a shock is, the larger the value of connectivity of the associated vertex is. Indeed, loops and multiple edges play a vital role in quantifying the strength of seismic activity of vertices. Therefore, a role of a "hub" is played by a main shock, and accordingly the preferential attachment rule may be satisfied, leading to a scale-free network [9].

In this paper, we analyze in the undirected network picture the hierarchical structure and mixing property of the earthquake network. We show that the clustering coefficient decays as a power law with respect to connectivity, manifesting hierarchical organization. This fact combined with the previous results obtained in Refs. [11-13] implies that there exist striking similarities between the earthquake network and the Internet. Then, we study the correlation property of the earthquake network by calculating the nearest-neighbor average connectivity and the Pearson correlation coefficient. We shall see that the network has assortative mixing. This point is of essential difference from the Internet with disassortative mixing. We present physical interpretations for these results in view of seismology.

We have constructed the earthquake networks in California and Japan by employing two different cell sizes, $10 \text{km} \times 10 \text{km} \times 10 \text{km}$ and $5 \text{km} \times 5 \text{km} \times 5 \text{km}$, which are considered to be reasonable if the typical size of a small fault and the resolution of measurement are taken into account. Following the procedure mentioned above, we have mapped to the growing random graphs two sets of the data made available by the Southern California Earthquake Data Center (http://www.data.scec.org/) covering the

region 28°36.00'N–38°59.76'N latitude and 112°42.00'W–123°37.41'W longitude with the maximal depth 175.99km in the period between 00:25:8.58 on January 1, 1984 and 22:50:49.29 on December 31, 2004, and by the National Research Institute for Earth Science and Disaster Prevention (http://www.hinet.bosai.go.jp/) covering the region 17°57.36N–47°59.88N latitude and 120°10.50E–154°29.64E longitude with the maximal depth 681.0km in the period between 00:02:29.62 on June 3, 2002 and 23:55:26.98 on March 31, 2005. The total numbers of the events are 379728 and 382639, respectively.

As shown in Refs. [11-13], these earthquake networks are small-world and scale-free. To investigate their hierarchical structure, first we analyze the clustering coefficient as a function of connectivity. This quantity is given as follow. Consider $c_i = 2e_i / k_i(k_i - 1)$, where $e_i$ is given by $e_i = (A^3)_{ii}$ with the adjacency matrix $A = (a_{ij})$ of a simple graph [that is, $a_{ij} = 1 (0)$ if the vertices $i$ and $j$ are connected (unconnected) and $a_{ii} = 0$] and $k_i$ is the value of connectivity of the $i$th vertex. Then, the clustering coefficient, $\bar{c}(k)$, is defined by $\bar{c}(k) = (1/[N P_{sg}(k)]) \sum_{i=1}^{N} c_i \delta_{k_i k}$, where $P_{sg}(k)$ stands for the connectivity distribution of the simple graph. This quantity gives information on hierarchical organization of the network. It is noticed that, upon calculating the clustering coefficient, loops have to be removed and multiple edges are to be replaced by simple edges in order to reduce the full network to the corresponding simple graph.

In Figs. 1 and 2, the plots of $\bar{c}(k)$ are presented. In both California and Japan, the clustering coefficient asymptotically decays as a power law. More precisely, it is nicely fitted by the $q$-exponential function over the whole range:

$$\bar{c}(k) \sim e_q(-k/\kappa), \tag{1}$$

where $e_q(x) = [1+(1-q)x]_+^{1/(1-q)}$ with $[a]_+ = \max\{0,a\}$, $q$ and $\kappa$ are positive constants, and, in particular, $q > 1$ in the present case. The inverse of the $q$-exponential function is the $q$-logarithmic function, which is given by $\ln_q(x) = (x^{1-q}-1)/(1-q)$. (In the limit $q \to 1$, the $q$-exponential and $q$-logarithmic functions converge to the ordinary exponential and logarithmic functions, respectively. These functions play central roles in nonextensive statistics [5].)

This hierarchical structure is of physical importance. The earthquake network has growth with preferential attachment [11], as in the Barabási-Albert model [9]. It is known [16] however that the Barabási-Albert model does not yield hierarchical organization. To mediate between growth with preferential attachment and the hierarchical structure, the authors of Ref. [17] have employed the concept of vertex deactivation [18]. This has a natural physical implication in the case of the earthquake network. Each fault may be deactivated through the process of stress release, in general. Furthermore, it should also be mentioned that the fitness model [19] can also generate hierarchical organization. It is our opinion that, in reality, the hierarchical structure of the earthquake network may be due to both deactivation and fitness, where the latter can be a function of magnitude.

We have also examined the dependence of the clustering coefficient on the threshold for the value of magnitude, $M_{th}$, from 0 to 3, and have found that the $q$-exponential behavior of $\bar{c}(k)$ disappears and no significant trends become observed already at $M_{th} = 2$. This implies that the hierarchical organization is mainly supported by weak shocks.

The above discovery (together with the scale-free nature realized by growth and preferential attachment reported in Ref. [11]) indicates the existence of striking similarities between the earthquake network and the Internet. To examine such similarities further, we present in Table I the time evolutions of some characteristic quantities of the earthquake network in California with the cell size $10\,\text{km} \times 10\,\text{km} \times 10\,\text{km}$. These results should be compared with those of the Internet [19-21].

However, there actually exists an essential difference between the earthquake network and the Internet. It is concerned with the mixing property. Therefore, finally we discuss the nearest-neighbor average connectivity in order to clarify the mixing property of the earthquake network. This quantity is given as follows. Consider the conditional probability, $P(k'|k)$, that a vertex of connectivity $k$ is linked to a vertex of connectivity $k'$. Then, the nearest-neighbor average connectivity of vertices with connectivity $k$ is defined by [19-21]: $\bar{k}_{nn}(k) = \sum_{k'} k' P(k'|k)$. In contrast to the preceding analysis of the clustering coefficient, here loops and multiple edges should be taken into account in order to quantitatively describe seismicity. In Figs. 3 and 4, the plots of this quantity are presented for California and Japan, respectively. In both cases, the feature of assortative mixing [21,22] is observed. That is, vertices with large values of connectivity tend to be linked to each other. It is also noticed that the Barabási-Albert model has no mixing [22]. Also, we have ascertained that, in contrast to the clustering coefficient, this trend of the nearest-neighbor average connectivity of vertices is robust against the threshold of magnitude at least up to $M_{th} = 3$.

To quantify such correlation, we have also calculated the Pearson coefficient [22]. This quantity is defined as follows. Let $e_{kl} (= e_{lk})$ be the joint probability distribution for an edge to be with a vertex with connectivity $k$ at one end and a vertex with connectivity $l$ at the other. Its marginal, $q_k = \sum_l e_{kl}$, obeys the normalization condition, $\sum_k q_k = 1$. Then, the Pearson correlation coefficient is given by $r = (1/\sigma_q^2) \sum_{k,l} kl (e_{kl} - q_k q_l)$, where $\sigma_q^2 = \sum_k k^2 q_k - (\sum_k k q_k)^2$ is the variance of $q_k$. $r \in [-1, 1]$, and $r$ is positive (negative) for assortative (disassortative) mixing.

The result is presented in Table II. Consistently with the one obtained from the analysis of the nearest-neighbor average connectivity, the Pearson correlation coefficient is positive, confirming that the earthquake network has assortative mixing.

On the other hand, the Internet is of disassortative mixing [19-22]. That is, the mixing properties of the earthquake network and the Internet are opposite to each other. (We have also ascertained that the presence of loops and multiple edges is essential for assortative mixing: the simple graph obtained by reducing the full earthquake network turns out to have disassortative mixing.)

In conclusion, we have found that the earthquake network exhibits hierarchical organization and therefore is not of the Barabási-Albert type. We have interpreted this fact in terms of vertex fitness and deactivation by the process of stress release at the faults. We have also found that the earthquake network possesses the property of assortative mixing. This point is an essential difference of the earthquake network from the Internet with disassortative mixing. Thus, correlation between the betweenness and degree centralities [10] is particularly strong for large values of connectivity. This may have importance for possible indirect prediction of main shocks by identifying the betweenness centralities.

The present work was supported in part by the Grant-in-Aid for Scientific Research of Japan Society for the Promotion of Science.

# Figure and Table Captions

**FIG. 1.** The log-log plots of the clustering coefficient with respect to connectivity

(insets: the semi-$q$-log plots) of the networks in California with the cell sizes

(a) $10 \text{km} \times 10 \text{km} \times 10 \text{km}$ and (b) $5 \text{km} \times 5 \text{km} \times 5 \text{km}$. The values of $q$ and $\kappa$ in the description in Eq. (1) are (a) $q = 1.18$, $\kappa = 5.38 \times 10^2$ and (b) $q = 1.77$, $\kappa = 5.46 \times 10^2$. All quantities are dimensionless.

**FIG. 2.** The log-log plots of the clustering coefficient with respect to connectivity (insets: the semi-$q$-log plots) of the networks in Japan with the cell sizes (a) $10 \text{km} \times 10 \text{km} \times 10 \text{km}$ and (b) $5 \text{km} \times 5 \text{km} \times 5 \text{km}$. The values of $q$ and $\kappa$ in the description in Eq. (1) are (a) $q = 1.89$, $\kappa = 3.45 \times 10^2$ and (b) $q = 1.98$, $\kappa = 3.85 \times 10^2$. All quantities are dimensionless.

**FIG. 3.** The log-log plots of the nearest-neighbor average connectivity of vertices with respect to connectivity of the networks in California with the cell sizes (a) $10 \text{km} \times 10 \text{km} \times 10 \text{km}$ and (b) $5 \text{km} \times 5 \text{km} \times 5 \text{km}$. The solid lines show the trends depicted by the exponentially increasing functions. All quantities are dimensionless.

**FIG. 4.** The log-log plots of the nearest-neighbor average connectivity of vertices with respect to connectivity of the networks in Japan with the cell sizes (a) $10 \text{km} \times 10 \text{km} \times 10 \text{km}$ and (b) $5 \text{km} \times 5 \text{km} \times 5 \text{km}$. The solid lines show the trends depicted by the exponentially increasing functions. All quantities

are dimensionless.

**TABLE I.**  Time evolution of the network in California with the cell size 10km×10km×10km: number of vertices (*N*), number of edges (*E*), average clustering coefficient ($<c>$), and average path length ($<l>$). Among these quantities, only $<c>$ is calculated after removing loops and replacing multiple edges by simple edges.

**TABLE II.**  The values of the Pearson correlation coefficient of the networks in California and Japan with two different cell sizes, 10km×10km×10km and 5km×5km×5km.